\documentclass{elsart}

\journal{Physica A}

\usepackage[dvips]{graphicx}
\usepackage{amsmath}
\usepackage{amsfonts}

\usepackage[numbers,sort&compress]{natbib}
\providecommand{\av}[1]{\left\langle #1 \right\rangle}
\providecommand{\neu}[1]{n_{\mathrm{#1}}}

\providecommand{\negF}{\phi_{1}}
\providecommand{\Wmax}{W_{\mathrm{max}}}

\begin{document}

\begin{frontmatter}
\title{Adaptivity and `Per learning'}
\author{Joseph Rushton Wakeling}
\address{Institut de Physique Th\'eorique, Universit\'e de Fribourg, P\'erolles, CH-1700 Fribourg, Switzerland}
\address{Niels Bohr Institute, Blegdamsvej 17, DK-2100 Copenhagen \O, Denmark}
\ead{joe@nbi.dk}
\ead[url]{http://neuro.webdrake.net/}

\date{17 March 2004}

\begin{abstract}
One of the key points addressed by Per Bak in his models of brain function was that biological neural systems must be able not just to learn, but also to \emph{adapt}---to quickly change their behaviour in response to a changing environment.  I discuss this in the context of various simple learning rules and adaptive problems, centred around the Chialvo-Bak `minibrain' model [Neurosci. 90 (1999) 1137--1148].
\end{abstract}

\begin{keyword}
adaptive learning \sep neural networks \sep feedback mechanisms \sep biological learning
\PACS 07.05.Mh \sep 84.35.+i \sep 87.18.Sn \sep 87.19.La
\end{keyword}

\end{frontmatter}

\section{Introduction}

When attempting to model biological learning, what factors should we take into account, and what sort of problems should we expect our models to solve?  One of the things which Per Bak always emphasized was that it was not enough simply to learn one task fast: biological neural dynamics had to be able to \emph{adapt}, to \emph{un}learn patterns of behaviour that were no longer working and find new ones.  For example, the important early work on `reinforcement learning' by Barto and colleagues~\cite{BSA83,B85,MAJ91}, which produced much more biologically plausible learning rules than those previously considered, still foundered on this problem, with networks having to be completely reset in order to learn a new problem.

Particular progress in this regard was made by the work of Dmitris Stassinopoulos, in collaboration with Preben Alstr{\o}m~\cite{AS95} and Per himself~\cite{SB95}.  However, it was a few years later that an especially elegant model was developed by Per in collaboration with Dante Chialvo~\cite{CB99}.  This model, which I rather cheekily dubbed the `minibrain'\footnote{Much to my embarrassment, after Per and I published our collaboration using this name, Per told me that he had always called it the `Dante brain', and Dante called it `Per learning'.  Readers are invited to draw their own conclusions but should not read anything whatsoever into the title of the present paper\ldots ;-)}\cite{WB01}, addressed the problem of adaptation by assuming that learning was by only \emph{long-term synaptic depression} (LTD), the weakening of connections.  Synapses involved in bad decisions were suppressed, but only enough to render them inactive.  Meanwhile, because synapses were not strengthened or reinforced in any way---there was a complete absence of long-term potentiation (LTP)---the strengths of active connections remained barely greater than those of the inactive ones.  Thus, the network could easily switch to using a different set of connections if the need arose.  It was suggested that it was the absence of any strengthening of connections in the model that was the key to its adaptive ability.  In the present work I illustrate this by investigating the adaptive ability of the minibrain when simple forms of LTP are included, compared to the original model and the `selective punishment' extension later proposed~\cite{BC01}.

\section{The model}

For simplicity I consider here only the basic feedforward minibrain: 3 layers of neurons (`input', 'intermediary' and `output') of size $\neu{ip}$, $\neu{im}$ and $\neu{op}$ respectively.  Each neuron in the input layer has a one-way connection to every neuron in the intermediary layer, and similarly each intermediary neuron has a connection to every output.  Each connection is assigned a strength value, initially evenly distributed in the interval $[0,W]$ (here $W=1$).  Activity propagates according to \emph{extremal dynamics}: if we stimulate a neuron, the signal travels along the single strongest outgoing connection, and the neuron at the end of that connection then fires, and so on until an output neuron fires.

Should this output be incorrect, a negative feedback signal is sent to the system and the connections responsible are punished by having their strengths reduced by a random amount in the interval $[0,\delta]$ (here $\delta=1$).  Learning efficiency is measured by the total number $\negF$ of such signals required for complete learning.  Depending on a control parameter $\zeta = \neu{im}/(\neu{ip}\neu{op})$, two phases of behaviour are identifiable~\cite{W03}: for $\zeta < 1$ the network is in the \emph{disordered} phase where complete learning is impossible and $\av{\negF} = \infty$.  For $\zeta > 1$ complete learning becomes possible with $\av{\negF} \approx \neu{ip}\neu{op}$.  In the present work, $\neu{ip} = 10$, $\neu{im} = 100$ and $\neu{op} = 10$, with the $\neu{im}$ value picked to ensure the network is in the ordered phase so complete learning will always be possible.

\section{\label{sec:unl-pos}Unbounded potentiation}

The simplest manner of including LTP is symmetric with the negative feedback: in the event of a successful decision, the connections responsible can be rewarded by having their strengths increased by a random amount in the interval $[0,\nu]$.  How does this affect the network's adaptive ability?

\begin{figure}[t]
\begin{center}
\includegraphics*[width=0.49\textwidth]{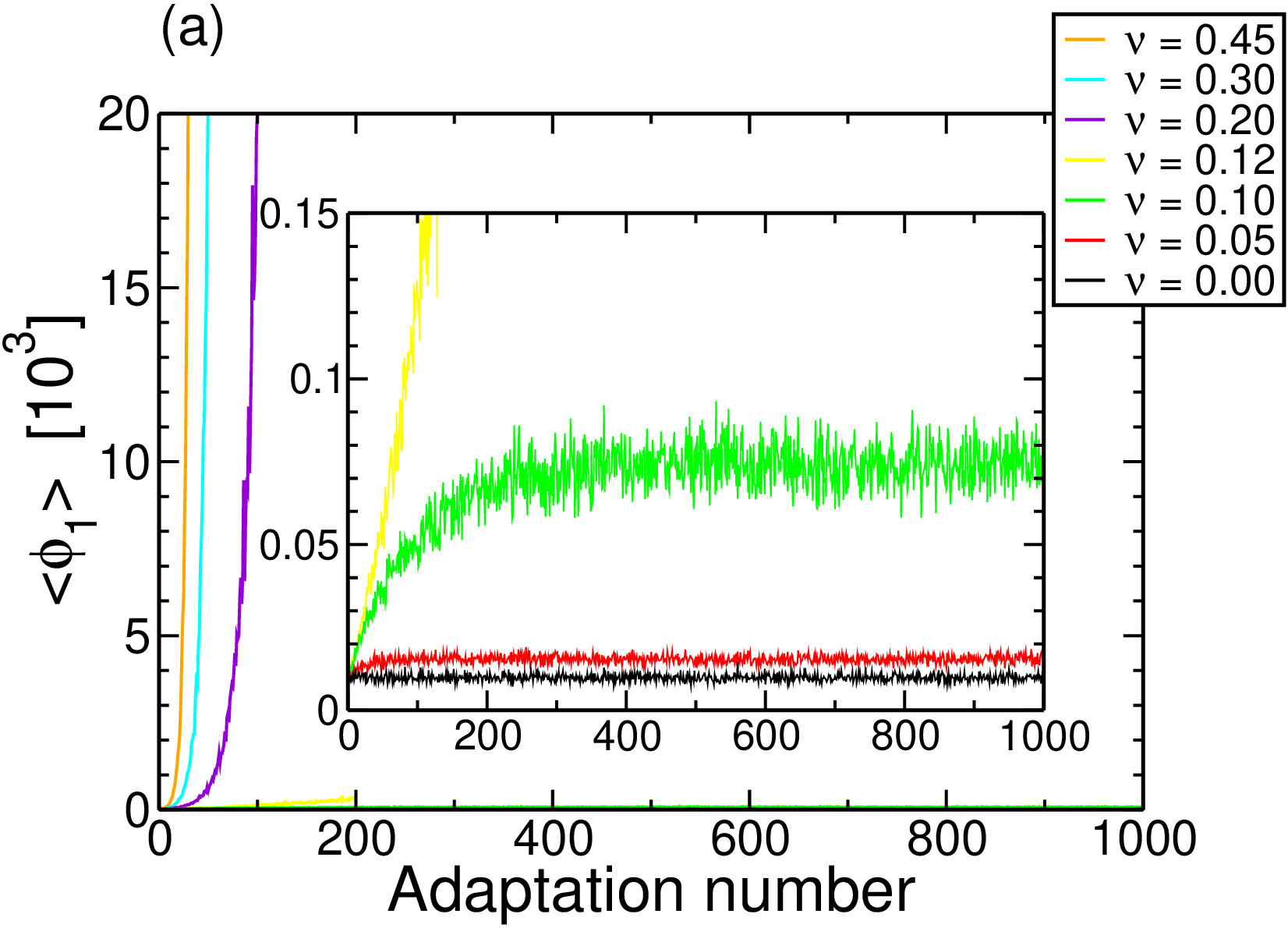}  \includegraphics*[width=0.49\textwidth]{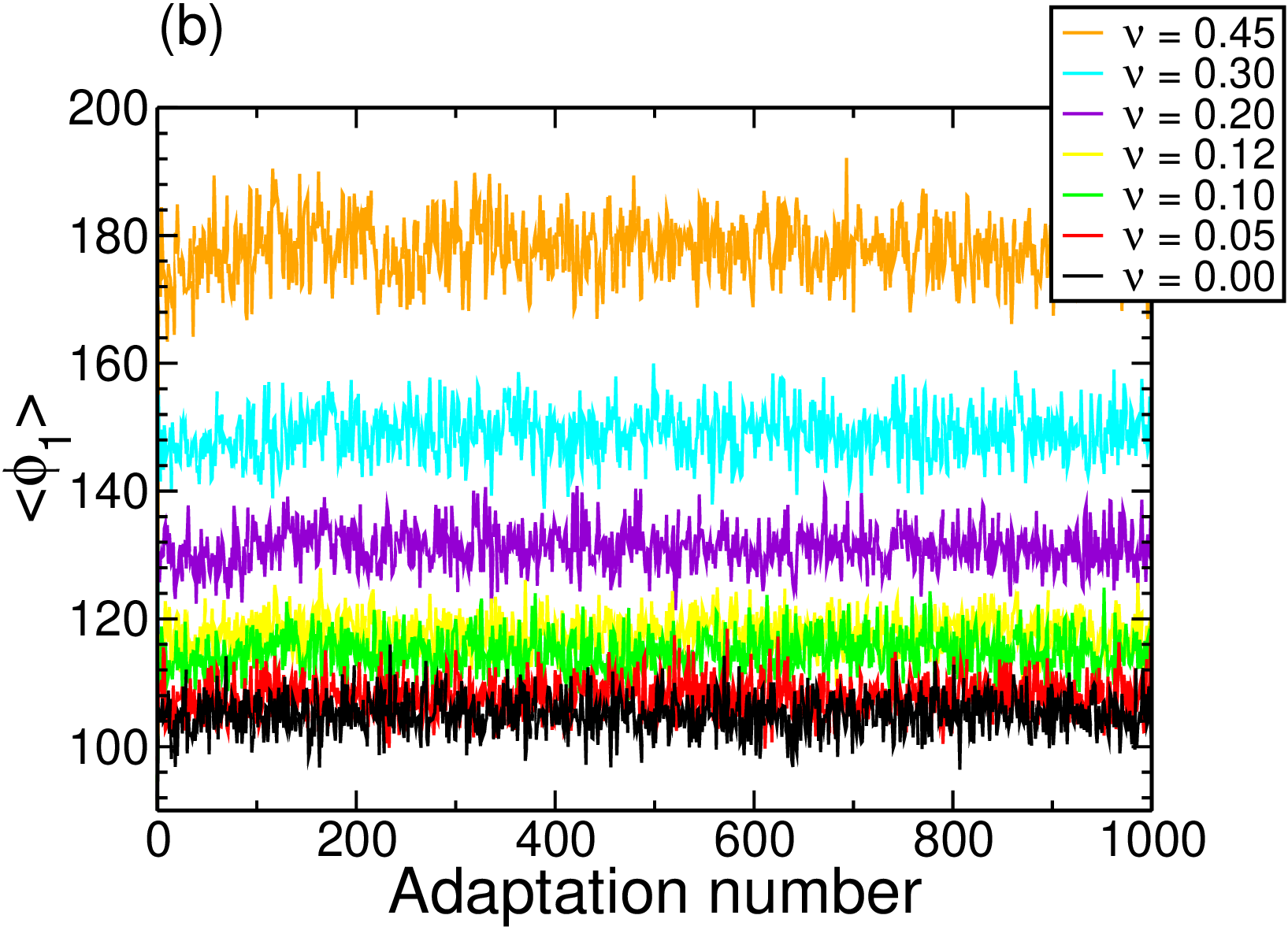}
\end{center}
\caption{\label{fig:ltd-ltp_unl}Adaptive ability of the minibrain model when unbounded LTP is included, compared to the traditional version with LTD only.  (a) At each adaptation, one input-output association is randomly reselected (`slow change').  The inset shows more clearly the results for smaller $\nu$.  Note the astonishing increase in the required value of $\negF$ for $\nu>0$: only for the smallest values, $\nu\leq 0.1 = 1/\neu{ip}$, is this bounded, and then performance is still significantly worse than for the LTD-only network.  (b) The network is required to adapt successively back and forth between two different input-output maps (the `flip-flop' problem).  $\av{\negF}$ is bounded in all cases, but nevertheless $\nu=0$ (LTD only) provides the best performance.  Data averaged over 128 realizations.}
\end{figure}

Suppose that we present a network with an input-output map to learn and, each time learning is completed, randomly reroll one of the input-output associations.  We can call this the `slow change' problem.  Naturally we want to know what value of $\negF$ the network will require to adapt to the new maps\footnote{Including LTP in this way might lead one to question whether it is still appropriate to use this measure of learning efficiency.  In fact this is a non-issue: LTP can only be applied to active connections, and a connection only becomes active through the depression of connections stronger than itself.  Therefore, even with positive feedback, it is still entirely appropriate to use $\negF$, the number of applications of negative feedback required for complete learning, as a measure of learning efficiency.  LTP in this context is not so much a learning mechanism as an `anti-forgetting' mechanism.}.  We stimulate each of the inputs in turn and apply LTP or LTD as necessary; learning is deemed complete when we can run through all the inputs without error.  Fig.~\ref{fig:ltd-ltp_unl}a shows how the average $\av{\negF}$ varies with successive adaptations.  Most of the networks with $\nu > 0$ very quickly become hopelessly addicted, requiring huge amounts of negative feedback to adapt to new input-output maps; this is escaped only with the smallest values of $\nu>0$.  Indeed, we can observe two distinct phases of behaviour: for $\nu\leq0.10$, $\av{\negF}$ is bounded (though always worse than the $\nu=0$ case), while for $\nu > 0.10$ we have the real addiction with $\av{\negF}$ growing exponentially.

We can explain this as follows.  Consider one input neuron.  On average, $\neu{ip}$ adaptations will pass before its associated output is reselected.  Since it is receiving positive feedback all the while, the amount of potentiation given to its active outgoing connections will be proportional to $\neu{ip}$.  Thus, if the ratio $\nu/\delta$ is greater than $1/\neu{ip}$, there will be a divergence between the strengths of the active and inactive connections, leading to the observed addiction.  More generally, for any $\nu>0$, it is possible to think of a rate of change slow enough that addiction will result.

Fig.~\ref{fig:ltd-ltp_unl}b shows the results for a different problem, the `flip-flop' problem.  This time, the network starts by having to learn the map $1\to 1$, $2\to 2$, $3\to 3$, \ldots, and, once this has been learned, the map required is switched to $1\to\neu{ip}$, $2\to(\neu{ip}-1)$, $3\to(\neu{ip}-2)$, \ldots; and we continue switching back and forth between these two inverse input-output maps.  Again, the learning process consists of repeatedly running through the cycle of inputs until the complete cycle can be run through without error.  The slowness of change of the input-output map is no longer an issue, since \emph{all} of the input-output mappings are changed at each adaptation.  Thus, as one should expect, the values of $\av{\negF}$ required to adapt remain bounded for all values of $\nu$.  Nevertheless, performance is still observably worse in all cases than the $\nu=0$ case with LTD only.

\begin{figure}[t]
\begin{center}
\includegraphics*[width=0.6\textwidth]{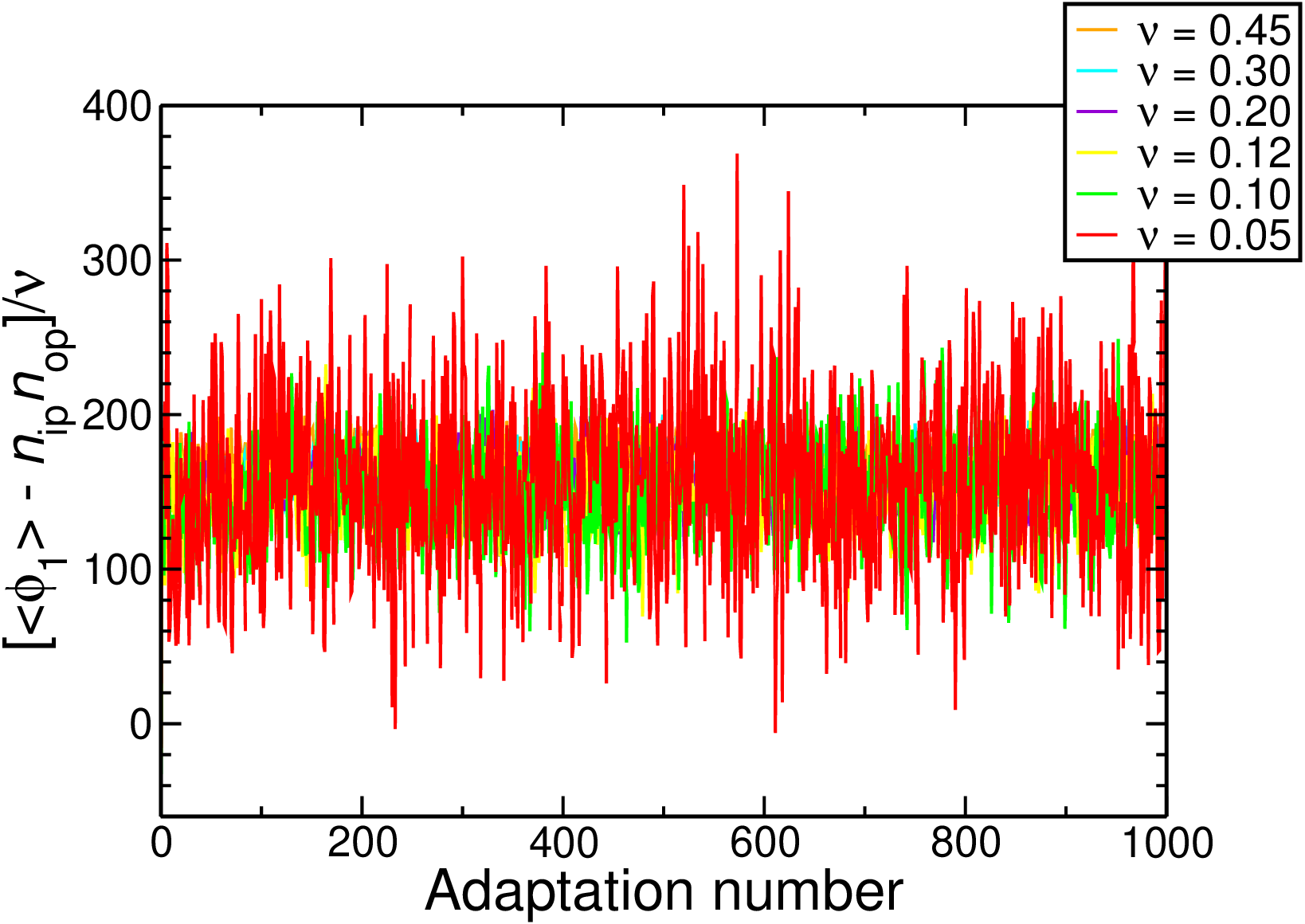}
\end{center}
\caption{\label{fig:ltd-ltp_ff-collapse}Collapse of data in Fig.~\ref{fig:ltd-ltp_unl}b for the flip-flop problem and unbounded LTP, according to the prediction of Equation~(\ref{eq:ff-unl}).}
\end{figure}

To explain this, consider again a single input.  Once it has been correctly wired up to its associated output, it will have a window of time (while the rest of the network is still learning) in which its active outgoing synapses will be potentiated.  The average for this time window will be controlled by the system size, i.e. a constant $f(\neu{ip},\neu{op})$ for any given network, and the total divergence between active and inactive connections will be proportional to $\nu$; thus the gain in the amount of negative feedback required to adapt will be given by $\nu/\delta$.  Mathematically speaking, we have,
\begin{equation}
\label{eq:ff-unl}
\av{\negF} = \neu{ip}\neu{op} + \frac{\nu f(\neu{ip},\neu{op})}{\delta}
\end{equation}
which is confirmed by the data collapse achieved in Fig.~\ref{fig:ltd-ltp_ff-collapse}.

\section{Bounded potentiation}

A method to avoid addiction while maintaining potentiation was proposed by Parisi~\cite{P86} with respect to the Hopfield neural network model.  By placing an upper bound on synaptic strength, he was able to construct a `memory which forgets' and thus avoid the state of total confusion observed if the network were overloaded.  This can be easily applied to the minibrain model, requiring synaptic strengths to be bounded in the interval $(-\infty,\Wmax]$, with $\Wmax\geq W$.  In this bounded case, should potentiation cause a synaptic weight $w_{ij}$ to go over the limit, we simply reset it to $\Wmax$.

\begin{figure}[t]
\begin{center}
\includegraphics*[width=0.49\textwidth]{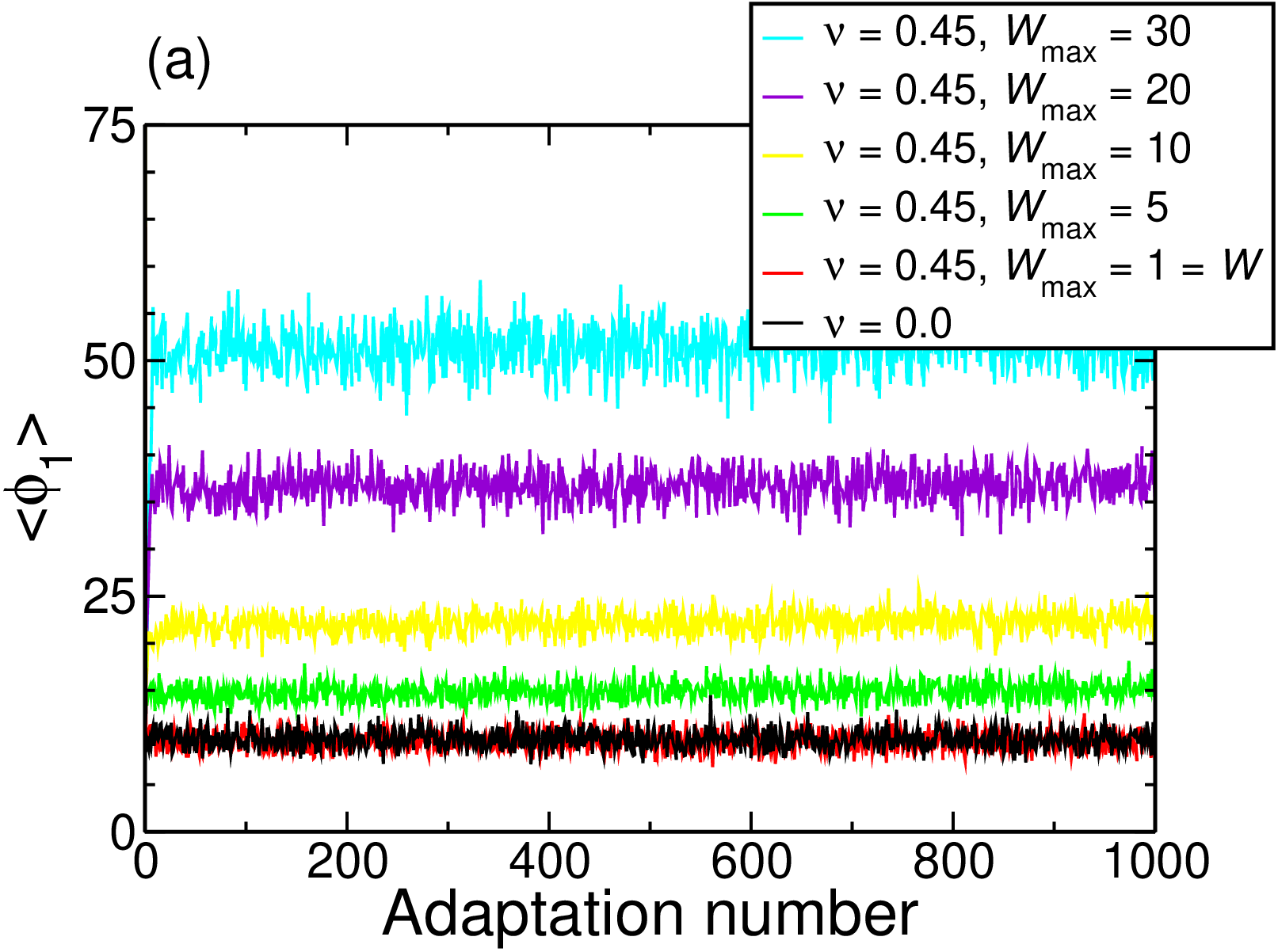}  \includegraphics*[width=0.49\textwidth]{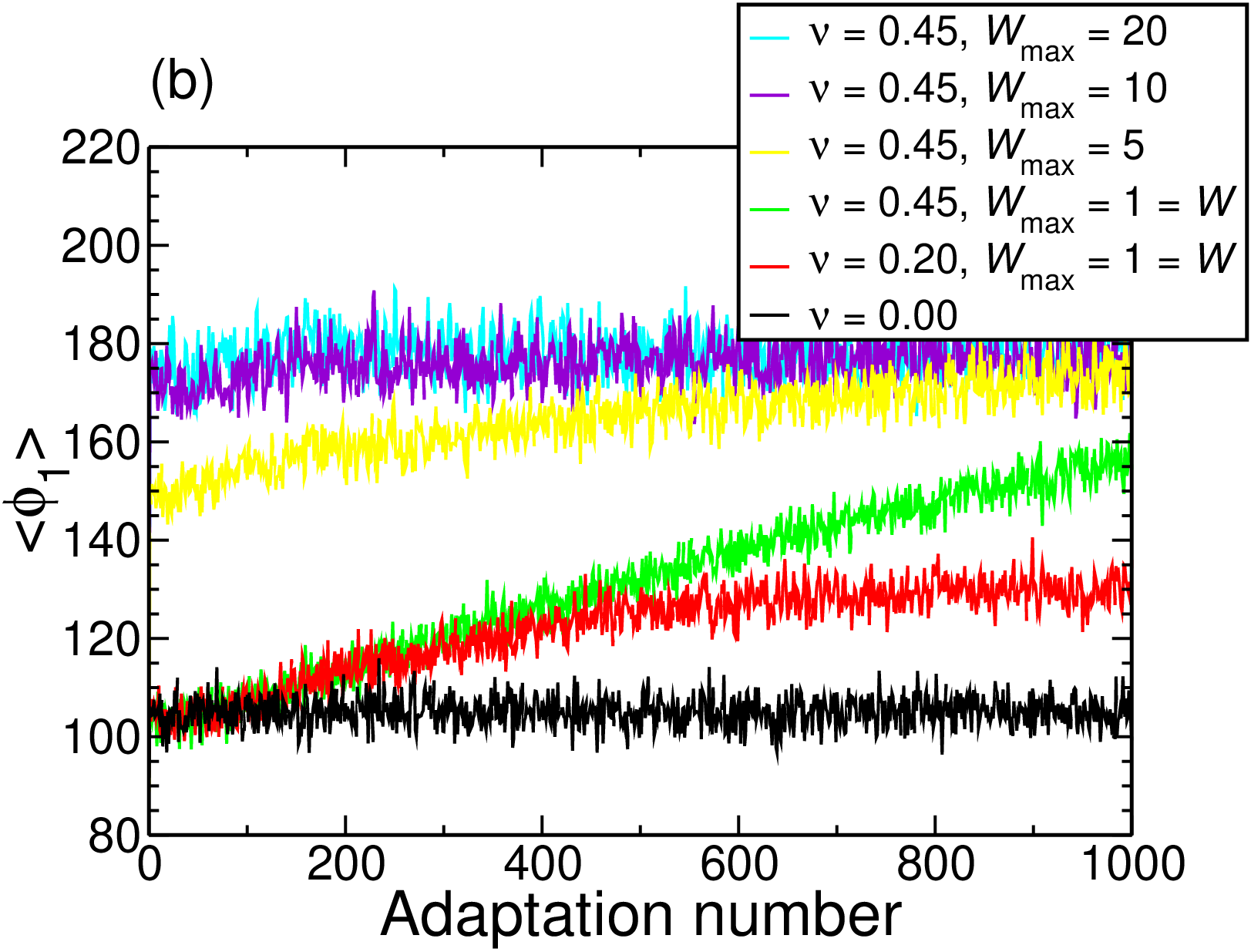}
\end{center}
\caption{\label{fig:ltd-ltp_cap}Adaptive ability when bounded LTP is included.  (a) Slow change problem.  Addiction is prevented but for $\Wmax > 1 = W$, adaptive ability is worse than in the LTD-only ($\nu=0$) version.  (b) Flip-flop problem.  $\av{\negF}$ increases to a maximum value determined by $\nu$ (see Fig.~\ref{fig:ltd-ltp_unl}b).  Data averaged over 128 realizations.}
\end{figure}

Fig.~\ref{fig:ltd-ltp_cap}a shows the performance in the slow change problem of different networks with $\nu=0.45$ and varying values of $\Wmax$, as compared to a network with $\nu=0$.  While the presence of the bound $\Wmax$ has prevented the runaway addiction seen in Fig.~\ref{fig:ltd-ltp_unl}a, there is no improvement over the simple LTD-only case.  In general, the network performs worse, and it is only with $\Wmax=W$ that the network matches the performance of the standard LTD-only minibrain.  This is natural when one thinks that inactive connections will always have strength of $o(W)$ whereas active connections will have strength $o(\Wmax)$.

By contrast a curious behaviour is observed in the `flip-flop' problem of switching back and forth between two different (non-overlapping) input-output maps (Fig.~\ref{fig:ltd-ltp_cap}b).  In all cases the value of $\av{\negF}$ increases with successive adaptations to a maximum value controlled, not by $\Wmax$, but by $\nu$ (see also Fig.~\ref{fig:ltd-ltp_unl}b).  How can we explain this?  Each time the input-output map switches, the system must suppress the active connections and then search among all connections for the required correct outputs.  What makes this different from the slow change problem is that here the majority of possible input-output connections are \emph{always} incorrect.  Therefore, the majority of connections will be continually weakened, never strengthened, and the gap between the average synapse strength and the maximum synapse strength $\Wmax$ will diverge.  This is equivalent to continually increasing the value of $\Wmax$, meaning that in the long run the system will behave as if this limit does not exist, reproducing the behaviour observed in the case of \emph{unbounded} LTP.

\section{Selective punishment}

Finally, let us consider the case of \emph{selective punishment}.  Here a synapse involved in a successful decision becomes permanently marked as `good'; should it later be punished, it is by an amount in the interval $[0,\delta^{\ast}]$ with $\delta^{\ast} < \delta$.  Thus, a previously good connection that has been suppressed is easier to reactivate than a connection that has never been good.

\begin{figure}[t]
\begin{center}
\includegraphics*[width=0.49\textwidth]{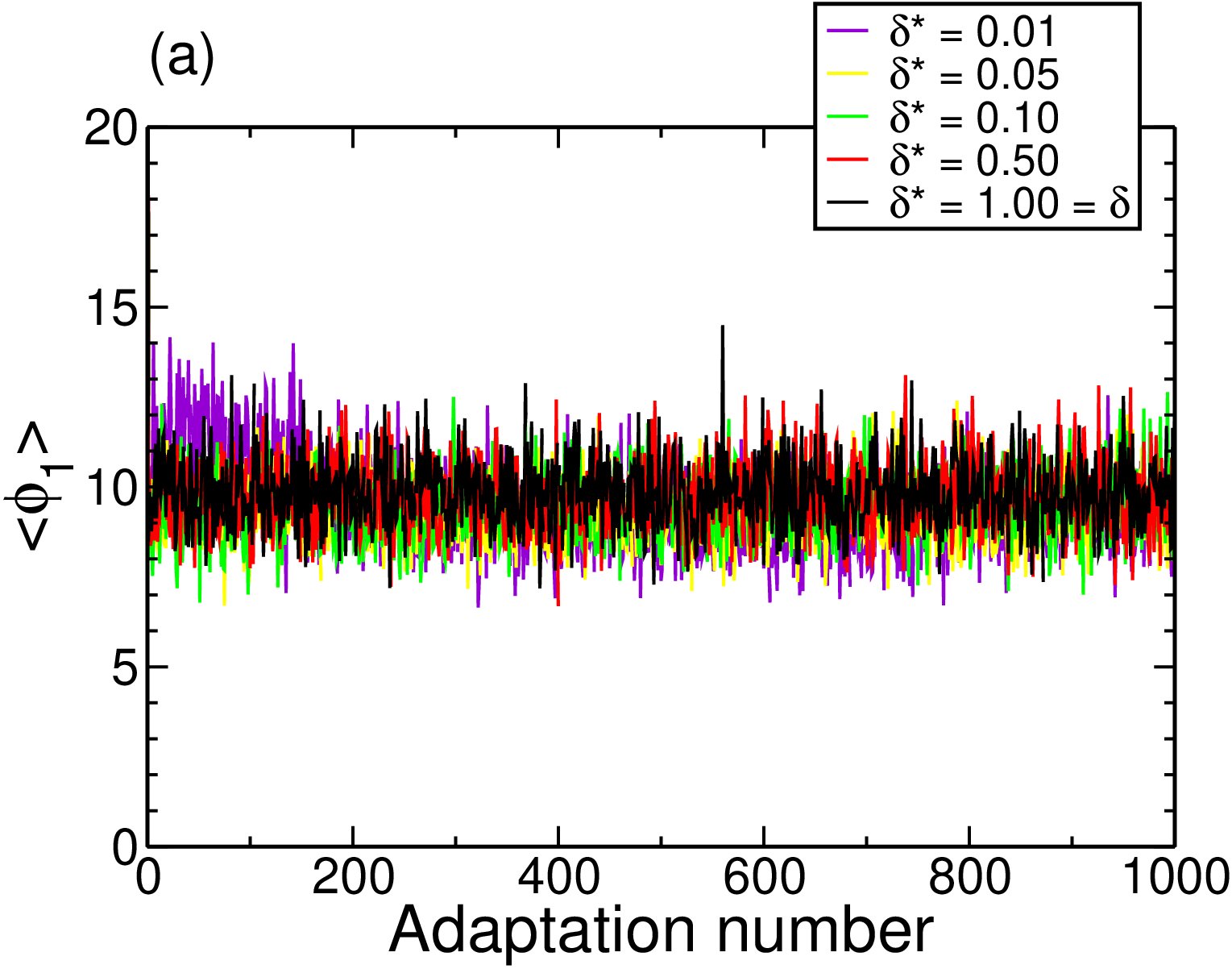}  \includegraphics*[width=0.49\textwidth]{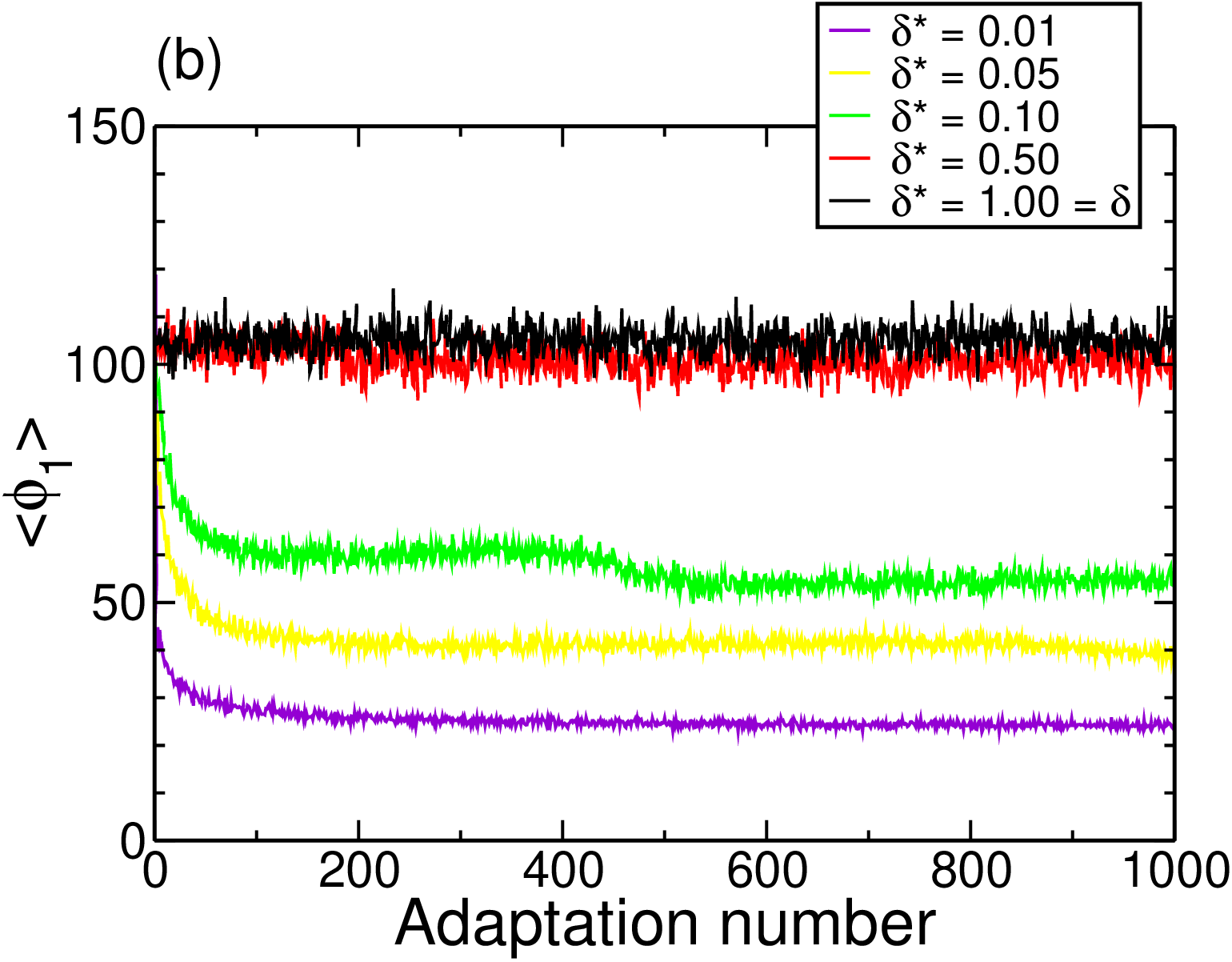}
\end{center}
\caption{\label{fig:ltd-ltp_sp}Adaptive ability with LTD only, but with selective punishment of previously successful connections.  (a) Slow change problem.  No significant difference is observed for different values of $\delta^{\ast}$.  (b) Flip-flop problem.  When lesser rates of punishment are applied to `good' connections, $\av{\negF}$ \emph{decreases} with successive adaptations: the system has a memory of previously good responses.  Data averaged over 128 realizations.}
\end{figure}

As Fig.~\ref{fig:ltd-ltp_sp}a shows, this makes no significant difference to the network's ability to adapt in the slow change problem\footnote{Different results are observed for different network topologies.  For example, on a random network, selective punishment proves very effective at enabling the system to distinguish between those paths that go nowhere or terminate in endless loops, and those that actually lead to output neurons~\cite{BC01}.}.  Should $\neu{im}$ take a smaller value, the effects of the selective punishment become more pronounced and adaptation is initially slower, but this effect vanishes as the network gains `good' connections to all possible outputs.

However, in the second of our two problems---switching back and forth between two distinct input-output maps---selective punishment proves a considerable benefit (Fig.~\ref{fig:ltd-ltp_sp}b).  While for $\delta^{\ast}=1.0=\delta$ the value of $\av{\negF}$ remains constant with a value of $\neu{ip}\neu{op}$, as one would expect, values of $\delta^{\ast}<\delta$ see a \emph{decrease} in $\av{\negF}$ with successive adaptations, towards a minimum value much lower than without the selective punishment.  Recall that the selective punishment favours previously-good connections over others.  Since in this case the number of good responses for each input is small by comparison to the total number of responses, this has the effect of drastically cutting learning times.

\section{\label{sec:concl}Conclusions}

Perhaps the key result of the present work has been to observe that, in the setup considered here, strengthening of synapses \emph{always} carries within it the potential for divergence between the strengths of active and inactive connections.  This divergence is governed not merely by the level of potentiation but also by the system size, increasing as the network becomes larger.

The minibrain is a `toy' model, but it is nevertheless instructive to consider it in the light of biological results.  Both LTP and LTD are well-observed in biological neuronal systems but their precise functions remain unclear \citep{S98,MGM00}.  A variety of different points of view can be found in the literature, with a number of authors explicitly endorsing a selectionist picture of neuronal dynamics~\cite{ST94} where learning is by either elimination or depression of connections.  Thinking along these lines one might want to seek other means of positive feedback than LTP, such as the `synaptic forgiveness' proposed by Klemm et al.~\cite{KBS00}.

Other authors have suggested that learning may result from a balance of LTP and LTD with a global feedback mechanism to prevent runaway strengthening or weakening of synapses \citep{RP03}.  A modification to the minibrain along these lines has recently been proposed by Bosman et al.~\cite{BvLW04}.  The present results suggest that such a global mechanism may not just be useful, but \emph{vital}, if LTP is to be an effective part of learning.

\section*{Acknowledgments}

Many thanks to Dante Chialvo and Maya Paczuski for helpful comments and advice, and for a great deal of personal and professional support over the last two years.  Thanks also to Yi-Cheng Zhang and Mogens H{\o}gh Jensen for making possible my sabbatical at the NBI where this work was completed.  I acknowledge the support of a grant from the Swiss National Science Foundation (no.~20-61470.00) and a Marie Curie Fellowship (HPMT-CT-2001-00402).

There are really no words adequate to express the debt of thanks and friendship I owe Per Bak, whose immense kindness and support, both professional and personal, made possible my whole scientific career.

\end{document}